\documentstyle[tighten,aps,epsf] {revtex}

\newcommand{\beq}{\begin{equation}}
\newcommand{\eeq}{\end{equation}}

\begin{document}

\draft

\title {A CTMC study of collisions \\
between protons and $H_2^+$ molecular ions}

\author{Fabio Sattin\thanks{E--mail: sattin@igi.pd.cnr.it}}
\address{Consorzio RFX and 
Istituto Nazionale per la Fisica della Materia, Unit\`a di Padova \\
Corso Stati Uniti 4, 35127 Padova, Italy}

\author{Luca Salasnich\thanks{E--mail: salasnich@hpmire.mi.infm.it}}
\address{Istituto Nazionale per la Fisica della Materia, Unit\`a di Milano \\
Dipartimento di Fisica, Universit\`a di Milano \\
Via Celoria 16, 20133 Milano, Italy}

\maketitle

\abstract{
We study numerically collisions between protons and $H_2^+$ molecular ions
at intermediate impact energies by using the Classical Trajectory 
Monte Carlo method (CTMC). Total and differential cross sections
are computed. The results are compared with: a) the standard 
one electron--two nucleon
scattering, and b) the quantum mechanical treatment of the $ H^{+} - 
H^{+}_{2} $ scattering.
}


\section{Introduction}

Ion--atom collisions represent one of the main fields of research in atomic 
physics, both experimental and theoretical.
Currently, there is a great deal of studies about the collisions with
electron transfer between ions and oriented atoms ({\it i.e.} with a
preferred sense of circulation of the electron around the nucleus)
and also between ions and aligned atoms (where the probability distribution
of the electron is not spherical).
The interest has been triggered by the development of experimental techniques
which allow to prepare atoms in well defined states of low 
\cite{houver} or high
quantum numbers $(n, l, m)$ \cite{macadam} and, after scattering, to measure the
final state of the system \cite{richter}.
The very narrow phase space volume sampled by the electron allows a
detailed study of the physical mechanisms occurring during the impact,
which would not be otherwise transparent due to the averaging over the
entire space of electronic configurations. Consequences
of first order effects--spatial overlap and velocity matching--have been
extensively studied with the above mentioned techniques, interpreted 
in the
light of classical mechanics \cite{lewar,homan}
 and sometimes compared with exact quantum calculations 
 \cite{dubois,lunds}.
A recent brief review about the latest developments in this field is given
by Schippers \cite{schippers}, and some even more recent experimental and theoretical
works can be found in \cite{lundy}. \\
Less attention has been paid to processes which involve more
than one target nucleus: some studies of collisions between $H_2$ and bare ions
are reported, for example, in \cite{schippers,deb,wang1,cheng,illesca}, 
and the scattering
$H^+$--$H_2^+$ is the subject of \cite{corchs}. \\
Calculations on these systems using quantum mechanics is far from 
being straightforward. By contrast, the application of classical 
methods, and in particular of the Classical Trajectory Monte Carlo 
method (CTMC) presents several advantages:
The numerical complications introduced by solving the equations of
motion for a few more particles are negligible. One may think, in comparison,
to the problems which arise when attempting of solving the Schr\"odinger
equation for three particles instead of two. \\
Since the original work of Abrines and Percival \cite{abrines}
the CTMC method has been
one of the most successful techniques
for studying the scattering between heavy charged particles at intermediate
impact energies. Starting from the simple $H$--$H^+$ processes and from the
calculation of simple total cross sections, the method has been refined so
that nowadays it gives many detailed informations:
Differential cross sections \cite{schultz}, which are ordinarily measured
in experiments, and state--to--state transitions \cite{salop}, 
which are especially useful for 
the research and development on thermonuclear fusion.
The CTMC method has also been applied to more complex systems:
Collisions involving more than an electron \cite{wetmore}, 
requiring the inter-electronic correlation, 
have been, and still are \cite{schultz,luca1}, a great challenge. 
\par
To our knowledge, until now CTMC has not been applied to polynuclear
targets: In this work we do an investigation on this kind of process
using the simplest target, the $H_2^+$ molecular ion. We aim to understand
how and to which degree the complex structure of the target modifies
the results of the scattering with respect to the standard two nucleons--one
electron case. \\
In presence of a diatomic target the question of interference between
the two scattering centres raises, and it has been faced 
within the quantum--mechanical formalism in Refs. 
\cite{deb,wang1,corchs}. In classical
calculations one cannot speak of interference as in quantum ones; however
a certain modification of the results due to the presence of a second
target is likely to be expected. 
\par
The first problem one has to face is to obtain an equilibrium electron
distribution function for the $H_2^+$ molecular ion.
In Section \ref{section2} it is described how this has been dealt with in 
this work.
\par
The results of the numerical simulations, expressed in the form of
cross sections, are compared with the similar results for two nucleons
scattering, and the differences are examined in Section \ref{section3}.

\section{Theory}
\label{section2}
In any CTMC calculation an important role is played by the choice
of the initial conditions of the electron in the phase space. 
It is well known that, in the case of the ground state of hydrogen, 
extracting electron coordinates from a microcanonical distribution, 
$ f({\bf r},{\bf p}) = \delta(E - E_{1s})/8 \pi^3 $,
yields the correct quantum mechanical momentum distribution [19]:
\beq
\tilde{\varrho}({\bf p}) = N { p^2 \over  (p^2 + 2 |E_{1s}|)^4} \quad ,
\eeq
where $ E_{1s} = -0.5 $ au (atomic units will be used unless otherwise
stated), and $N$ the normalization factor. Obviously this relations holds 
also for hydrogen--like ions. On the other hand, the radial distribution is not 
reproduced satisfactorily.
While no classical method can reproduce at the same time both the exact
momentum and spatial distribution, some methods have been devised
which allow to exactly reproduce the latter distribution at the 
expenses of the former (so called CTMC-{\it r} method, 
opposed to the CTMC-{\it p} method--see \cite{cohen} ), 
or to yield an approximate--but rather 
good--description of both distributions (for more about the subject see the
references quoted in \cite{schultz} or \cite{luca2}). 
In this work we have approximated the quantum mechanical electron wave
function by a Linear Combination of Atomic Orbitals:
\beq
\psi({\bf r}) = {\psi_{1s}^A({\bf r}) + \psi_{1s}^B({\bf r})
                 \over \sqrt{2} } \; ,
\eeq
where $A$ and $B$ refer to the two protons which are placed
initially with null velocity along the $z$ axis, 
at a distance $\pm z_0$ from the origin, with $z_0$ kept equal to 1 au,
in accordance with the true equilibrium internuclear distance.
With this  choice the
molecule has a cylindrical symmetry with respect to the $z$ axis. 
From a Fourier transform of Eq. (2) we get the momentum
distribution function: since 
\beq
\psi_{1s}^A({\bf r}) + \psi_{1s}^B({\bf r}) =
\psi_{1s}({\bf r} - {\bf r}_0) + \psi_{1s}({\bf r} + {\bf r}_0)
\quad ,
\eeq
where ${\bf r}_0=(0,0,z_0)$, we obtain
\beq
{\tilde \psi}({\bf p}) = {1 \over (2 \pi)^3} \int \; d{\bf r} \;
\psi({\bf r}) \; e^{ - i {\bf p} \cdot {\bf r} }
= {1 \over \sqrt{2} } \left( e^{ - i p_z z_0}
\tilde{\psi}_{1s}(p) + e^{ i p_z z_0} \tilde{\psi}_{1s}(p)
\right)  \quad .
\eeq
The probability density is better expressed in cylindrical coordinates
\beq
\tilde{\varrho}({\bf p}) = 2 \pi p_r |\tilde{\psi}({\bf p})|^2
= N' {p_r \over (1 + p_r^2 + p_z^2)^4} \cos^2(p_z z_0) \quad ,
\eeq
where $p_z , p_r$ are the projections of $\bf p$ along
the $z$ axis and the radial direction in the {\it x-y} plane.
$N'$ is a normalization factor.
\par
The probability density 
$\tilde{\varrho} $ is similar to a hydrogenic distribution (see Eq. 1)
but for the factor $ p_r \cos^2(p_z z_0)$.
This means that electron velocity is preferentially found within
the {\it x-y} plane, where $p_z \simeq 0$.
\par
The couples $p_r$, $p_z$ are picked up within a range $(0 , p_{max}) $
and generated according to the distribution of Eq. (5) with a rejection
technique; $p_{max}$ is chosen great enough so that
$\tilde{\varrho}(p_r,p_z)$ is negligible for $ p_r^2 + p_z^2 > p_{max}^2 $:
in the computations $p_{max} = 4.5$.
\par
The choice of the spatial position of the electron follows a similar
route. The position $\bf r $ is constrained to satisfy
\beq
{p^2 \over 2 m_e} - { 1 \over | {\bf r} - {\bf r}_0 |} -
{ 1 \over | {\bf r} + {\bf r}_0 |}  = E_{H_2^+} \quad ,
\eeq
where ${\bf r}_0 = (0 , 0 , z_0)$ and $E_{H_2^+} \simeq -1.1$ au 
is the experimental quantity.  $\bf r$ is characterized
by the three numbers: the radial distance $ r = |{\bf r}|$,
the polar angle $\theta$ and the azimuthal angle $\phi$.
Again, $ r $ and $\cos(\theta)$ are chosen with a rejection technique
within the ranges $ ( 0, r_{max} )$ and $( -1 , 1)$ respectively.
$\phi$ does not explicitly appear in Eq. (6) 
and may be chosen uniformly in the interval $(0, 2 \pi )$. \\
The
chosen distribution is not expected to be a stationary one for the classical
system, so it is of interest to see how much it varies with time.
In Figure 1 we plotted the contour of the distributions 
$\tilde{\varrho}(p_z, p_r)$ $\varrho(z,r)$ at $ t = 0$ and $t = 5 $, 
in absence of the projectile,
from which one may see that the difference is not small but the essential 
features still remain so the choice of this distribution appears to 
be justified.
\par
The projectile initial parameters are the velocity $v$, the impact
parameter $b$ with respect of the centre of mass $O$ of the molecule,
the azimuthal angle $\vartheta$ with respect to the molecule axis, and
the initial distance $d$ from $O$. A sketch of the scattering configuration
for the coplanar case (all the nuclei lying on the same plane)
is shown in Figure 2. The projectile is then rotated out of the plane by
a random angle between 0 and $2 \pi$.
\par
After having initialized all the four particles, the corresponding
equations of the motions are numerically integrated in time until the
nucleons are well far apart. In the computations $d$ is not a critical
parameter as far as it is great enough to allow the target molecule and
the projectile to be considered initially as non--interacting. After
some trials a value of $d \simeq 20 $ au was found to be reasonable.
With given values for $ b$, $v$ and $\vartheta$, a number $N$ of runs have
been carried on, varying only the electron initial conditions.
At the end of each collision one may find one of the following situations:
1) the original molecule remains intact; 2) it may be dissociated but the
electron is still bound to one of the two nuclei; 3) the electron is bound
to the projectile (we call this case "charge transfer''); 4)
the molecule may be broken and the electron be ionized.
Each process $i$ (with $i=1,2,3,4$) happens $N_i$ times over
the total $N$ runs, with corresponding probabilities
\beq
 P_i (v, b, \vartheta)= {N_i \over N }  \quad .
\eeq
The standard deviation error of $P_i$ is
\beq
\Delta P_i = \sqrt{{ N - N_i \over N N_i}} P_i \quad .
\eeq
Cross sections are computed by integrating over the impact parameter $b$
\beq
\sigma_i (v, \vartheta) = 2 \pi \int_{0}^{b_{max}} \; db \; b\; P_i(b) \quad .
\eeq
where $ P_i(b) \simeq 0$ for $ b > b_{max} $.
Still, the cross section of Eq. (9) may be averaged over $\vartheta$:
\beq
\tilde{\sigma }_i (v) = {1 \over 2 \pi} \int_{0}^{2 \pi} d\vartheta \;
  \sigma_i (v , \vartheta) \quad .
\eeq
This latter integral has been evaluated by a simple trapezoidal rule
using the values  $\vartheta_k$ (k=1,...,n) for a finite set of angles.

\section{Results}
\label{section3}
The runs have been performed for $ 0.3 < v < 2.0  $ au.
The choice is done to include the region of maximum 
effectiveness of the CTMC method: $ v \ge v_e$, with $v_e$ electron velocity.

The presence of a second nucleus is clearly seen when one plots $ \sigma$
{\it versus} $ \vartheta$ (Figure 3). One can see an increasing trend 
with $ \vartheta$, {\it i.e.} electron capture is favoured when 
the projectile impinges with a direction perpendicular to the molecule
axis. $\vartheta$ plays here the same role of the angle $\phi$ between the
angular momentum of an aligned electron and the projectile
direction in ion--atom collisions (see, for example, Figure 1 of Ref. 
\cite{lunds}):
with this parallelism in mind, the data may be compared with similar plots,
for example, in \cite{homan,lunds,wang2}. 
In comparison with those cases the effect is here much less marked, due to the
fact that the electron probability distribution is smeared over a broader
phase space volume. Nevertheless it seems possible to
give at least a qualitative explanation of the trends in Figure 3 using
propensity rules. As already explained in Section II one finds that,
for a given value of the momentum $p$, the maximum of the
probability of finding a matching between the velocities of the electron and
the projectile is when $p_z = 0 $ and--as $ p_z = p \cos{(\vartheta)}
$--this means $\vartheta = \pi /2$ (see also Figure 1, where the electron
distribution is localized close to $ p_z = 0$).  
$\sigma (\vartheta = 90^\circ)$ is almost constant up to $ v \simeq 1$
and only then falls down.
In ref. \cite{wang1} similar plots have been obtained for the scattering
$ H^+$--$H_2 $ at high velocities ( $ \geq 1$ MeV), within the
Brinkman--Kramers formalism. There is discernible (see their Figure 6)
a fluctuation, attributed to interference effects, which does not appear
in our data (this was to be expected since, obviously, purely quantum
mechanical  effects cannot be included in our model). The same effect, even
enhanced, is experimentally found in \cite{cheng}.
\par
One way of looking at these data is plotting the anisotropy parameter
\beq
 A(v) = {\sigma_{cx}(v , \vartheta = 0^\circ) - 
         \sigma_{cx}(v, \vartheta = 90^\circ) \over
       \sigma_{cx}(v, \vartheta = 0^\circ) + 
        \sigma_{cx}(v, \vartheta = 90^\circ) }
\eeq
{\it versus } $v$ (Figure 4). $\sigma_{cx}$ is $\sigma_i$ from Eq.
(10) corresponding to the process of electron capture.
$A$ is oscillating but definitely assumes negative value, approaching zero
while $v$ increases. $A < 0$ 
means that capture is favoured when the projectile and the molecular
alignment are orthogonal. This is in agreement with other works 
(see, for example, the paper by Thomsen {\it et al} or 
that of Olson and Hoekstra in ref. \cite{lundy}) where, furthermore, 
a more complex behaviour is also found, with changes of sign of $A$. 
\par
In Figure 5(a) total cross sections (the same data of Figure 3,
averaged over angle $\vartheta$)
are shown as functions of impact velocity $v$. These data lend themselves
to a comparison with structureless target scattering:
in Janev \cite{janev} it is empirically demonstrated how nucleus--hydrogen
scattering follows a scaling law: the curve
$ \sigma_{cx} / n^4 Z $ {\it versus} $ v^2 n^2 / Z^{0.5}$ is universal,
regardless of the initial principal quantum number $n$ of the electron
and of the charge $Z$ of the nucleus.
We may imagine to replace the diatomic molecule with a single particle,
to which the electron is bound 
in a state defined by an effective (non integer) quantum number
$ n_{eff} = 1/ \sqrt{2 |E_{H_2^+}|} \simeq 0.67 $ with our
values. The agreement between our rescaled data and the universal 
curve by Janev yields an estimate of how much this modelling is
justified. From Figure 5(b) one sees that the qualitative trend is the same,
and the data are quite well interpolated by the fit in the middle
of the range $v$: it is expected that, with increasing $v$, the electron--ion
collisions closer and closer resemble two body processes, with a lesser
influence of the target nucleus. In this situation the distribution function
should not have influence. At the lower $v$'s, the suggested fit
underestimates the data; however, it is difficult to discerne how much of this
discrepancy is due to the structure of the target and how much to the
intrinsic defects of the CTMC method in this region of low energy.
\par
In order to have a further insight about the reliability of our results,
we have compared them with previous calculations performed with other
methods: in ref. \cite{corchs} a calculation similar to ours has been carried on
in the impact energy range from 100 keV to 5 MeV ($ 2 \leq v \leq 14 $)
using a distorted-wave model under different approximations: the simpler
OBK approximation and the more refined correct-boundary-conditions Born
serie (B1B) and the first order Bates series (Ba1) (see \cite{corchs} and
references therein for more details about these approximations). Figure 4
of ref. \cite{corchs} shows the differential cross section for electron capture
$d\sigma / d (\cos \vartheta)$ as a function of $\vartheta$ at a collision
energy of 100 keV for $H^+$--$H^+_2$ collisions. 
We have integrated the curves plotted 
and the results are shown in Figure 5(a). From this one may see that the
accuracy of our calculation (at least for the single energy point
available) is of the same order as the OBK approximation, and therefore
overestimates the correct value, which should be close to that given by
the B1B and Ba1 methods (which better fit the Janev' scaling law). 
\par
Up to now, only $\tilde{\varrho}({\bf p})$ has been taken into account to
justify the results, so it is interesting to study the effects
due to the spatial distribution $\varrho({\bf r})$.
Looking at the differential cross section
${d\sigma /db} \propto P(b) b$
for various impact energies and azimuthal angles,
we noticed that the increase of $\sigma$ with $\vartheta$ is due to
the contribution from larger $b$'s. This agrees with the results of 
\cite{wang1}.
\par
Finally, some words about the final state distribution. In our system 
about $80 - 90 \%$ of the total captures occur in the ground state. 
This is easily justified because the 
electron prefers to preserve its energy before and after the capture. 
A detailed study, looking for example at a dependence of this 
distribution from azimuthal angle or energy, 
would need a much larger amount of data, beyond the possibilities 
of the present study. 

\section{Summary and conclusions}

A series of numerical simulations has been performed on the charge--transfer
collisions between protons and hydrogen molecular ions using classical
methods. The interest of the subject relies on the comparison between this
system and other, well studied, three-particle systems.
Some conclusions which may be drawn from this study are:
I) The CTMC method applied to this target is able to discerne its
structure--as is seen from differential cross sections--but, with respect
to quantal methods, its sensitivity is greatly reduced, as may be seen from
the fact that no fluctuations due to interference effects are seen;
II) Besides partial cross sections, also total cross sections seem to depend 
on the
structure of the target, but this point is more difficult to stress
since main differences appear at small $v$'s, where CTMC is less reliable;
III) The accuracy of the CTMC has been compared with quantal methods in the
region of high $v$, limiting to total cross sections. It is found
that--within the very small data set--the predictions of the CTMC well
agree with those of the less refined  versions of the quantum mechanical
calculations, and slightly overestimate the more refined ones.

\newpage

\begin{figure}
\epsfxsize=12cm
\epsfbox{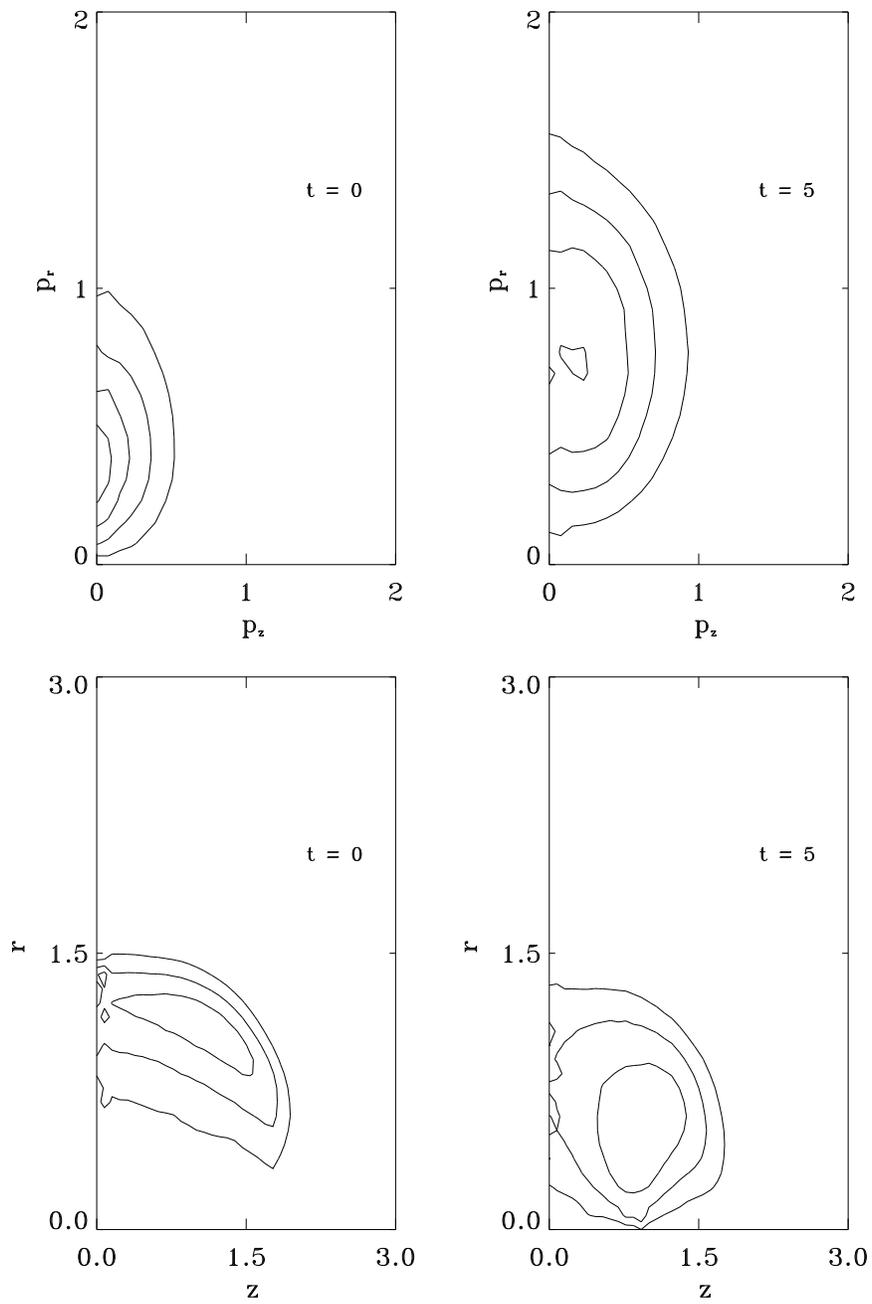}
\caption{
Contour plot of $\tilde{\varrho}(p_z,p_r)$ and of
${\varrho}(z,r)$ at $ t = 0 $ and $ t = 5 $.
}
\end{figure}

\newpage

\begin{figure}
\epsfbox{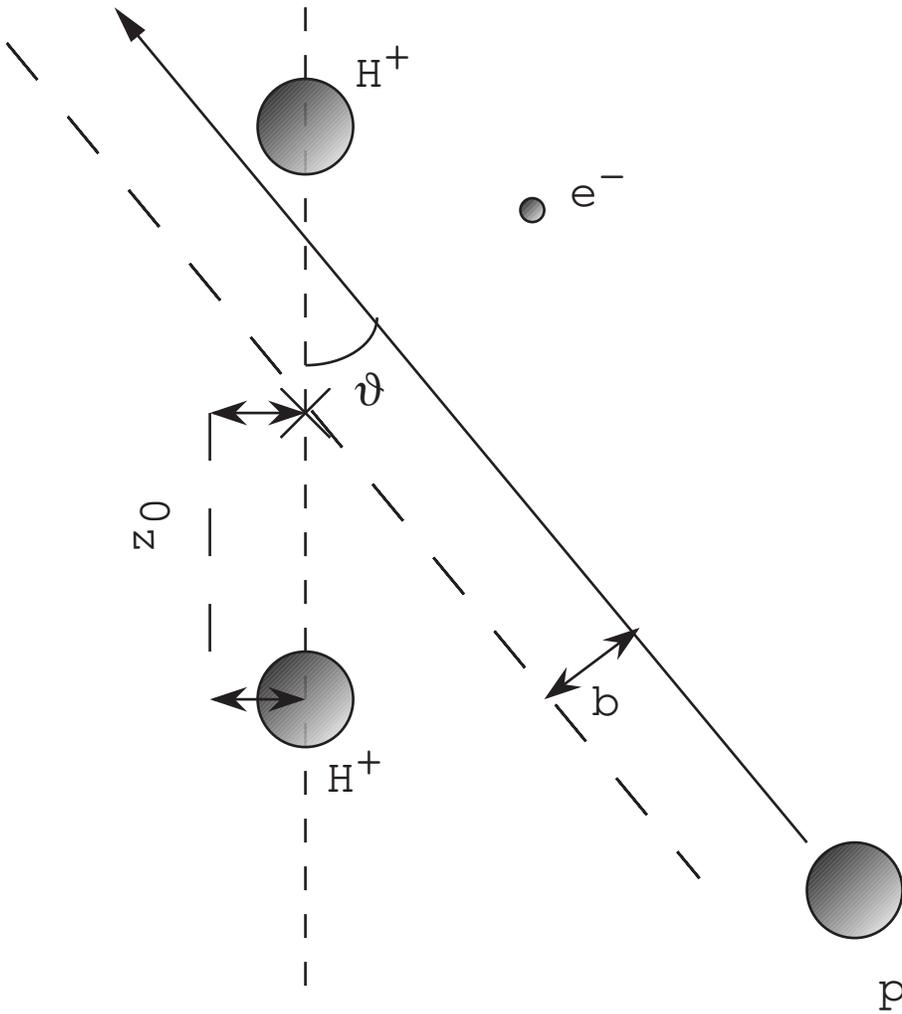}
\caption{
Sketch of the geometrical arrangement of the collision.
The two $H+$ are the nuclei of the $H^+_2$ molecule, $p$ is the projectile
and $e^-$ the electron. $z_0$ is half the internuclear distance, $b$ the
impact parameter, $\vartheta$ the angle of impact and $d$ the initial
distance. For easiness, a coplanar collision is sketched: in the general
case the projectile must be rotated out of the plane of an angle $\phi_{rot}$
uniformly chosen within the range ($0 , 2 \pi$). 
}
\end{figure}

\begin{figure}
\epsfxsize=13cm
\epsfbox{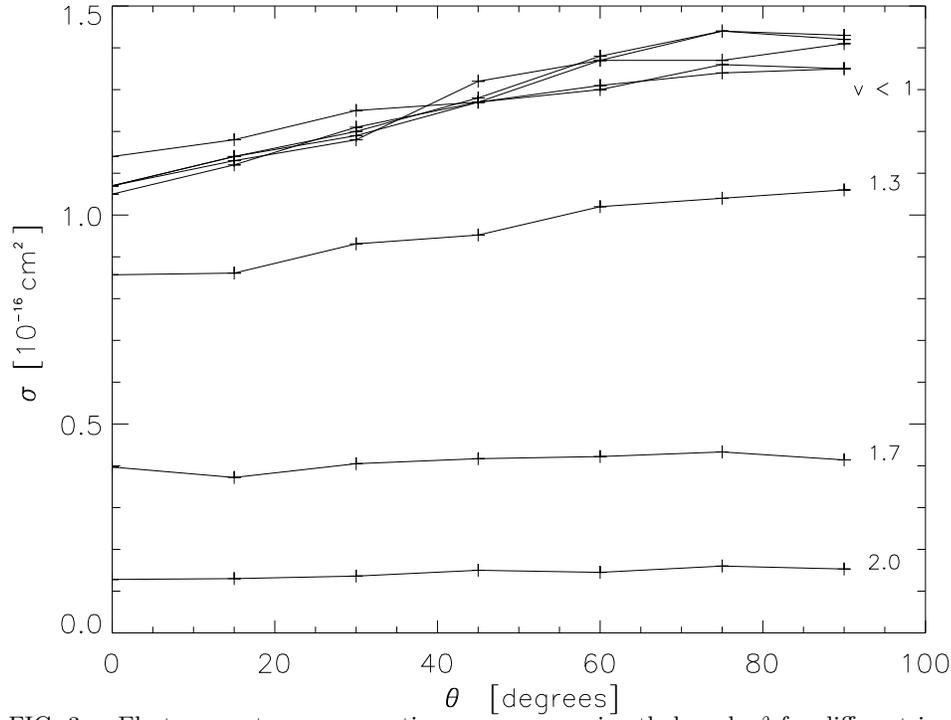}
\caption{
Electron capture cross section $\sigma_{cx}$ {\it versus}
azimuthal angle $\vartheta$ for different impact velocities. 
Errors bars are not shown as they are of the same size of the symbols. 
}
\end{figure}

\begin{figure}
\epsfxsize=13cm
\epsfbox{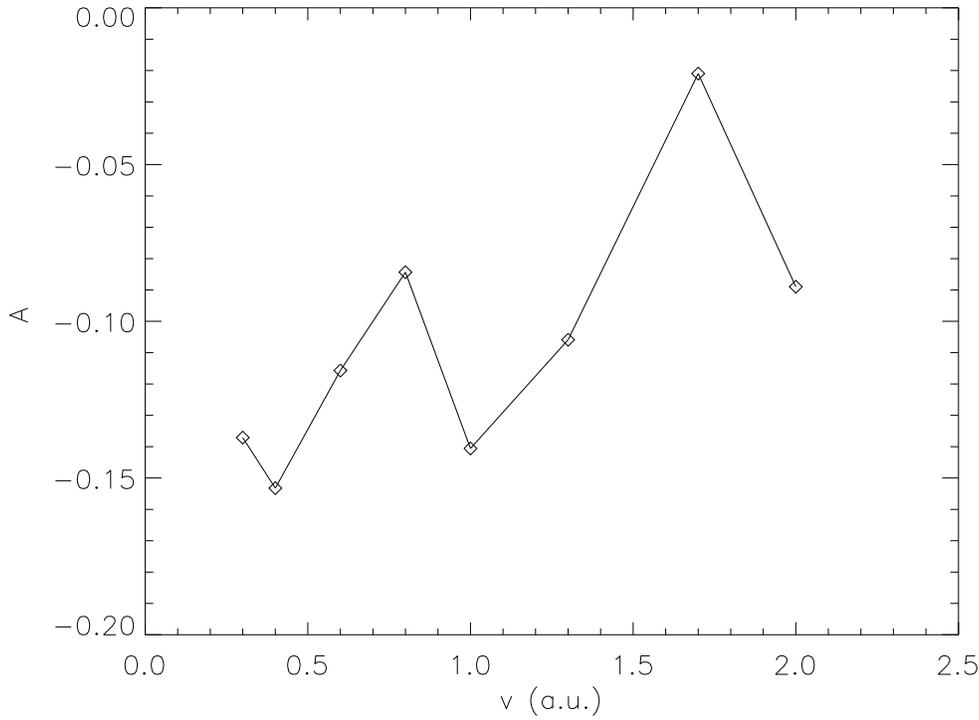}
\caption{
Anisotropy parameter $A$ {\it versus} $v$. 
}
\end{figure}

\begin{figure}
\epsfxsize=13cm
\epsfbox{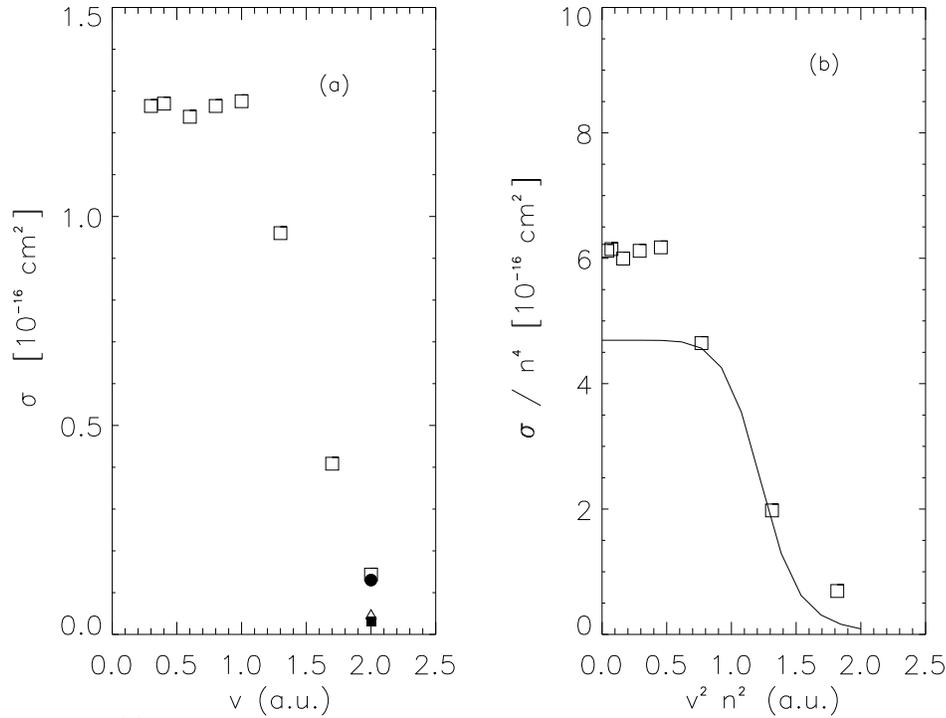}
\caption{
(a) Total electron capture cross section 
$ \sigma_{cx} $, averaged over angle $\vartheta$, {\it versus} $v$.
At $ v = 2$ are also shown the data taken from ref. [14]: 
OBK approximation (full circle), B1B approximation (full square),
Ba1 approximation (full triangle); see also the text. (b) The same data,
but rescaled according Janev [23]: $\sigma_{cx} / n_{eff}^4$
{\it versus} $v^2 n_{eff}^2$. 
The squares are the results from the present work, the solid line the fit
from ref. [23]. 
}
\end{figure}

\end{document}